\documentclass[preprint,1p,times,twocolumn,longmaketitle]{elsarticle}

\usepackage{graphicx} % Required for inserting images
\usepackage{amssymb}

\usepackage{natbib}

\usepackage{dcolumn}% Align table columns on decimal point
\usepackage{bm}% bold math
\usepackage{import} %imports other tex files
\usepackage{lineno} % line numbers
\usepackage{hyperref} %makes the references colored
\hypersetup{
    colorlinks=true,
    linkcolor=blue,
    filecolor=blue,      
    urlcolor=blue,
    citecolor=blue,} 
\usepackage[justification=raggedright]{caption}
\usepackage{subfig}

\date{\today}

\usepackage{lineno}
\makeatletter
\def\ps@pprintTitle{%
 \let\@oddhead\@empty
 \let\@evenhead\@empty
 \def\@oddfoot{}%
 \let\@evenfoot\@oddfoot}
\makeatother

\journal{Physics Letters B}

\begin{document}

\begin{frontmatter}

\title{First measurement of differential cross sections for muon neutrino charged current interactions on argon with a two-proton final state using the MicroBooNE detector}

% Authors in alphabetical order

\author[nn]{P.~Abratenko}
\author[nn]{O.~Alterkait}
\author[o]{D.~Andrade~Aldana}
\author[v]{L.~Arellano}
\author[mm]{J.~Asaadi}
\author[kk]{A.~Ashkenazi}
\author[l]{S.~Balasubramanian}
\author[l]{B.~Baller}
\author[cc]{A.~Barnard}
\author[cc]{G.~Barr}
% ...
%\author[]{\\[12pt] (The MicroBooNE Collaboration) \corref{eee}}
%\cortext[eee]{microboone\_info@fnal.gov}

% ...
\address[l]{Fermi National Accelerator Laboratory (FNAL), Batavia, IL 60510, USA}
% ...
\address[o]{Illinois Institute of Technology (IIT), Chicago, IL 60616, USA}
% ...
\address[v]{The University of Manchester, Manchester M13 9PL, United Kingdom}
% ...
\address[cc]{University of Oxford, Oxford OX1 3RH, United Kingdom}
% ...
\address[kk]{Tel Aviv University, Tel Aviv, Israel, 69978}
% ...
\address[mm]{University of Texas, Arlington, TX, 76019, USA}
\address[nn]{Tufts University, Medford, MA, 02155, USA}
% ...
%add     

\begin{abstract}
We present the first measurement of differential cross sections for charged-current muon neutrino interactions on argon with one muon, two protons, and no pions in the final state. These final states are dominated by two-nucleon knockout interactions, which are complicated to model and for which there is currently limited information about the characteristics of these interactions in existing neutrino-nucleus scattering data. Detailed investigations of two-nucleon knockout are vital to support upcoming experiments exploring the nature of the neutrino. Among the different kinematic quantities measured, the opening angle between the two protons, the angle between the total proton momentum and the muon, and the total transverse momentum of the final state system are most sensitive to the underlying physics processes as embodied in various theoretical models. 
\end{abstract}

%%\item Research highlight 1
%%\item Research highlight 2


%\maketitle

%\makeatletter
%\patchcmd{\ps@pprintTitle}{\footnotesize\itshape
%       Preprint submitted to \ifx\@journal\@empty %Elsevier
%       \else\@journal\fi\hfill\today}{\relax}{}{}
%\makeatother

\end{frontmatter}

\textit{*Email address:} \texttt{microboone\_info@fnal.gov}

%Introduction:f
%%%%%%%%%%%%%%%%%%%%%%%%%

\section{Introduction}
\label{introduction}

The introduction of the liquid argon time projection chamber (LArTPC) \cite{Rubbia,Chen:1976pp,WILLIS1974221,Nygren:1974nfi} has revolutionized the field of accelerator-based neutrino physics, allowing more detailed observations of ionizing radiation emitted from final state particles than was previously possible. This development has highlighted the need for advanced modeling of neutrino-nucleus interactions, especially neutrino-argon interactions, and detailed measurements to benchmark those models. Cross-section measurements for various nuclei and final state topologies are needed to support the development of neutrino interaction models \cite{modeling}. These models must address both in-medium nuclear modification of the fundamental neutrino interactions and also final-state interactions (FSI) involving the reaction products as they exit the nucleus~\cite{ALVAREZRUSO20181}. \\ 
\indent One process that probes both neutrino interactions and nuclear effects is the production of two-particle two-hole (2p2h) states in which two nucleons are removed from the nucleus. These states are primarily produced by neutrino interactions where the momentum transfer is shared between two nucleons via the exchange of a virtual meson, known as a meson exchange current (MEC)~\cite{ALVAREZRUSO20181}. In addition, 2p2h states can be produced by nuclear effects, such as short-range nucleon-nucleon correlations (SRC)~\cite{src_general,src} and FSI. In the case of SRCs, the neutrino interacts with a nucleon that is part of a correlated nucleon-nucleon pair. The momentum is transferred to a single nucleon but, because this nucleon is part of a correlated pair, both nucleons are knocked out of the nucleus. In the case of FSI, it is possible for a single nucleon to knock out a second nucleon as it exits the nucleus thereby leading to a 2p2h final state. Observation of 2p2h states in electron scattering has been used to develop the models of 2p2h production in neutrino scattering \cite{RUIZSIMO2018323,PhysRevD.101.033003}.  Correct modeling of 2p2h interactions is of vital importance to neutrino energy reconstruction and precision measurements of neutrino oscillations. Difficulty in measuring this production mode has required large variations to models to agree with data \cite{Acero2020, uboone_tune} and has the potential to bias energy reconstruction if the prediction of the 2p2h production cross section is incorrect \cite{Abe2017,Abe2023}. %Critically, the production cross-section has never been measured directly in neutrino scattering.

\indent A final state topology consistent with the production of a 2p2h state is a charged-current (CC) muon neutrino ($\nu_\mu$) interaction that results in one muon, two protons, and no charged or neutral pions (CC1$\mu$2p0$\pi$). While there is an existing measurement of CC1$\mu$2p0$\pi$ events on argon, no cross sections were extracted \cite{argoneut}. 
%In detectors with a high threshold for proton reconstruction (see for example \cite{PhysRevD.98.032003}) the observation of 2 protons in the final state is limited by the small available phase space with 2 high-momentum protons, and thus this final state cannot be distinguished from other CCQE interactions. 
In this letter, we present the first differential cross section measurements of CC1$\mu$2p0$\pi$ topologies on argon using data collected from the Micro Booster Neutrino Experiment (MicroBooNE) \cite{Acciarri2017}.

\section{Detector and Samples}
\label{DetectorSamples}

The MicroBooNE experiment uses an 85 metric ton active volume LArTPC detector located at Fermi National Accelerator Laboratory~\cite{Acciarri2017}. The detector is situated on-axis to the Booster Neutrino Beam (BNB) which has an average energy of $\langle E_\nu \rangle = 0.8$ GeV \cite{bnb} and is located approximately 470\,m from the neutrino production target. The detector consists of two components: a TPC, 10.36\,m long in the beam direction, 2.56\,m wide in the drift direction, and 2.32\,m tall; and an optical system comprised of 32 eight-inch photomultiplier tubes (PMTs). The TPC consists of three wire planes, two induction planes and one collection plane. The collection plane is oriented vertically and the induction plane wires are oriented at angles $\pm60^{\circ}$ with respect to the vertical direction.
%, with the PMTs mounted in an array behind the collection plane. 
%Both components are enclosed in a cylindrical cryostat containing 170 metric tons of liquid argon with approximately 85 metric tons within the TPC volume. 
Electronic signals from the TPC wire planes and the PMTs are recorded and subdivided into two distinct data samples. The first sample, known as on-beam data (BNB data), is collected coincident to a 1.6\,$\mu s$ BNB neutrino spill. The second sample, known as off-beam data (EXT data), is recorded in anti-coincidence with the beam. 
%The purpose of this second sample is to account for electronic noise and the large cosmic muon background caused by the surface location of the MicroBooNE detector. 
This sample provides a measure of the electronic noise and large cosmic muon background caused by the surface location of the MicroBooNE detector. 
The BNB data and a portion of the EXT data samples are then filtered based on a required minimum amount of activity measured in the PMTs. This study uses data from the three-year period 2016-2018, corresponding to $6.85\times 10^{20}$ protons on target.\\ 
\indent MicroBooNE utilizes the GENIE neutrino event generator \cite{genie} to create two samples of simulated neutrino interactions: Monte Carlo which addresses signal events and beam-related backgrounds (Beam MC), and Monte Carlo which addresses backgrounds from neutrino interactions in the material surrounding the cryostat (Dirt MC). 
The simulated events from both samples are then combined with events from an unbiased EXT sample (which is not filtered based on PMT activity) in order to simulate MicroBooNE's cosmic ray background \cite{uboone_cosmic_rejection}. Furthermore, MicroBooNE utilizes another sample of simulated events generated using the NuWro neutrino event generator \cite{nuwro} for additional studies.  \\ 
\indent The neutrino-argon scattering model in both the Beam MC and Dirt MC samples is given by the GENIE MicroBooNE Tune, a version of the GENIE v3.0.6 G18\_10a\_02\_11a model set in which four parameters are tuned to  $\nu_\mu$ CC0$\pi$ data from the T2K experiment \cite{uboone_tune}. The four parameters are: the CC quasi-elastic axial mass \cite{GENIE_v3}, the strength of the random phase approximation (RPA) corrections in the Nieves CCQE cross section calculation \cite{nieves_rpa}, the absolute normalization of the CC2p2h cross section \cite{nieves_mec}, and the shape of the CC2p2h cross-section. The shape of the CC2p2h cross section is represented by a parameter that linearly interpolates between two models: the Valencia prediction \cite{nieves_mec} and the Dytman model \cite{mec}.

\section{Event Selection}
\label{EventSelection}

The signal consists of CC $\nu_\mu$ interactions within the fiducial volume with the final state particles consisting of exactly one muon with momentum 0.1 $\leq$ $P_{\mu}^\mathrm{true}$ $\leq$ 1.2\,GeV/$c$ and two protons with momentum 0.3 $\leq$ $P_{p}^\mathrm{true}$ $\leq$ 1.0\,GeV/$c$. Events with any number of neutral pions are excluded. Signal events may also contain protons with momentum below 0.3\,GeV/$c$ or above 1.2\,GeV/$c$, any number of neutrons, and charged pions with momentum below 65\,MeV/$c$.

%The signal consists of charged-current muon-neutrino interactions within the fiducial volume with the final state particles consisting of exactly one muon with momentum 0.1 $\leq$ P$_{\mu}$ $\leq$ 1.2\,GeV/$c$, two protons with momentum 0.3 $\leq$ P$_{p}$ $\leq$ 1.0\,GeV/$c$, no charged pions with momentum above 65\,MeV/$c$, no neutral pions, and any number of neutrons.

The BNB data, EXT data, Beam MC, and Dirt MC samples are processed by the Pandora reconstruction framework \cite{pandora} to identify and reconstruct tracks from the ionized signal. The track's energy deposition and length are used to reconstruct momentum and particle identification. From the reconstructed products, a series of selection requirements are applied to identify CC1$\mu$2p0$\pi$ events. The initial selection retains events that meet three criteria: (1) the reconstructed neutrino vertex is located within a fiducial volume (FV) defined to be at least 10\,cm inside any TPC face, (2) there are exactly three tracks, and no showers as determined by Pandora, (3) the three tracks start within 4\,cm of the reconstructed neutrino vertex. Calorimetry-based particle identification techniques described in Ref.~\cite{pid} are then used to identify events with a single muon candidate and two proton candidates. \\ 
\indent  The final event selection requires one muon with momentum in the range 0.1 $\leq$ $P_{\mu}^\mathrm{reco}$ $\leq$ 1.2\,GeV/$c$ and two protons, both with momentum in the range 0.3 $\leq$ $P_{p}^\mathrm{reco}$ $\leq$ 1.0\,GeV/$c$, to match the signal definition.  The limits on momentum are driven by the resolution effects as well as phase space regions which have non-zero efficiencies and well-understood systematic uncertainties.  To benchmark the performance of the selection, the number of simulated CC1$\mu$2p0$\pi$ events that pass all selection requirements is compared to the number of events generated using the GENIE MicroBooNE tune; the event selection described above achieves an efficiency and purity of $\simeq$$14\%$ and $\simeq$$65\%$, respectively. 
%More details of the event selection can be found in Ref.~\cite{my_thesis}.

Several theoretical predictions of neutrino interactions are compared to data within this work and a detailed summary of each prediction is provided in the supplemental materials. Four GENIE model \cite{GENIE_v3} sets are considered: GENIE v3.0.6 G18\_02a\_00\_000 (GENIE Empirical), GENIE v3.0.6 G18\_10a\_02\_11a (GENIE Nieves), GENIE v3.2.0 G21\_11b\_00\_000 (GENIE SuSAv2), and the GENIE MicroBooNE Tune. Each of these GENIE models uses different models for QE, MEC, and FSI. The GENIE Empirical model set uses the relativistic Fermi gas (RFG) nuclear model \cite{RFG}, the Llewellyn Smith QE model \cite{LlewellynSmith}, and an empirically derived prediction for MEC interactions \cite{mec}. The GENIE Nieves model set uses the local Fermi gas (LFG) nuclear model \cite{LFG}, which is similar to the RFG model but includes considerations for the nuclear density. The Nieves QE \cite{Nieves_QE} model is similar to the Llewellyn Smith QE model, but includes contributions from long-range nucleon-nucleon correlations and Coulombic effects. The Nieves MEC model is based on a macroscopic calculation \cite{nieves_mec}. The SuSAv2 model set utilizes the relativistic mean field (RMF) approximation nuclear model \cite{RMFA,RMFA_includes_mesons}, which considers relativistic effects in the calculation of the motion of the nucleons in the nucleus. Relativistic effects are also considered in the calculation of the SuSAv2 QE \cite{PhysRevD.101.033003} and SuSAv2 MEC \cite{susa_mec} predictions by using scaling functions, derived from relativistic assumptions, to scale the cross sections at high momentum transfers. %Although GENIE lacks the option to use an RMF nuclear model directly, it achieves approximate consistency with the RMF-based results by choosing the nucleon initial momentum from an LFG distribution. In addition, the default nucleon binding energy used in GENIE for the LFG model is replaced for SuSAv2 with an effective value tuned to most closely duplicate the RMF distribution~\cite{PhysRevD.101.033003,PhysRevD.103.113003}. 
The GENIE MicroBooNE Tune is identical to the GENIE Nieves model except that two CCQE and two CC2p2h parameters are tuned to T2K data \cite{uboone_tune}, as described previously.

Beyond the GENIE generator predictions, the GiBUU 2023 \cite{Mosel2024}, NuWro 19.02 \cite{Golan2012}, and NEUT 5.4.0 \cite{Hayato2021} model sets are also considered. The GiBUU model set uses an LFG model \cite{LFG}, a standard CCQE expression \cite{Leitner2006}, and an empirical MEC model. These models are consistently implemented to solve the Boltzmann-Uehling-Uhlenbeck transport equation \cite{Mosel2019}. The NuWro model set is built on the same LFG model \cite{Carrasco1992}, using the Llewellyn Smith QE model \cite{LlewellynSmith}. The FSI treatment in NuWro uses an intranuclear cascade \cite{Nieves2011} to transport the hadrons through the nucleus, along with a coupling to \texttt{PYTHIA} for hadronization \cite{Sjstrand2006}. The NEUT model set utilizes an LFG model \cite{Carrasco1992}, with Nieves CCQE \cite{Nieves_QE} and MEC \cite{nieves_mec} scattering models, and treats FSI with medium corrections for pions \cite{genie}.

Using samples of simulated $\nu_\mu$ CC interactions from each model set, a study was conducted to identify variables sensitive to physics differences between the model sets. One such variable is the opening angle between the protons in the lab frame, 
$\theta_{\vec{P}_L \cdot \vec{P}_R}$,
defined as:
\begin{equation}\label{eq:opening}
     \cos(\theta_{\vec{P}_L \cdot \vec{P}_R}) = \frac{\vec{P}_L \cdot \vec{P}_R}{\left|\vec{P}_L \right| \, \left|\vec{P}_R\right|}
\end{equation}
where $\vec{P}_L$ is the momentum of the leading proton, the proton with the highest momentum, and $\vec{P}_R$ is the momentum of the recoil proton, the second proton in the event. A second physics-sensitive variable is the opening angle between the muon and total proton momentum vector, 
$\theta_{\vec{P}_{\mu} \cdot \vec{P}_{\mathrm{sum}}}$, defined as: 
\begin{equation}\label{eq:opening_muon}
    \cos(\theta_{\vec{P}_{\mu} \cdot \vec{P}_{\mathrm{sum}}}) = \frac{\vec{P}_\mu \cdot \vec{P}_\mathrm{sum}}{\left|\vec{P}_\mu \right| \, \left|\vec{P}_\mathrm{sum}\right|}
\end{equation}
where $\vec{P}_\mu$ is the momentum of the muon and $\vec{P}_\mathrm{sum}$ is the vector addition of the leading and recoil proton momenta. 
The opening angle between the protons in the lab frame provides information on the effect of the QE and MEC modeling on the proton momentum, while the opening angle between the muon and total proton momentum vectors provides information on the treatment of the outgoing muon momentum in relation to the 2p2h system. 

In addition to these two angles, we also identify the magnitude of the momentum transverse to the neutrino beam direction of the final state system, $\delta P_T$~\cite{Lu:2015tcr}. The transverse momentum vector of the CC1$\mu$2p0$\pi$ system ($\delta \vec{P}_T$) is defined as:
\begin{equation} \label{eq:delta_PT}
    \delta \vec{P}_T = \vec{P}^{\mu}_T + \vec{P}^L_T + \vec{P}^R_T
\end{equation}
where $\vec{P}^{\mu}_T$, $\vec{P}^L_T$, and $\vec{P}^R_T$ are the transverse momentum vectors of the muon, leading proton, and recoil proton respectively. 
%The magnitude $\delta P_T$ is sensitive to nuclear effects, with $\delta P_T \ne 0$ indicative of motion between the nucleons in the initial state.
The magnitude $\delta P_T$ is sensitive to nuclear effects, final state interactions, or below threshold undetected particles.

\begin{figure}[!h]
\captionsetup[subfigure]{labelformat=empty}  
\centering
\subfloat[][(a)]{{\includegraphics[width=0.8\columnwidth]{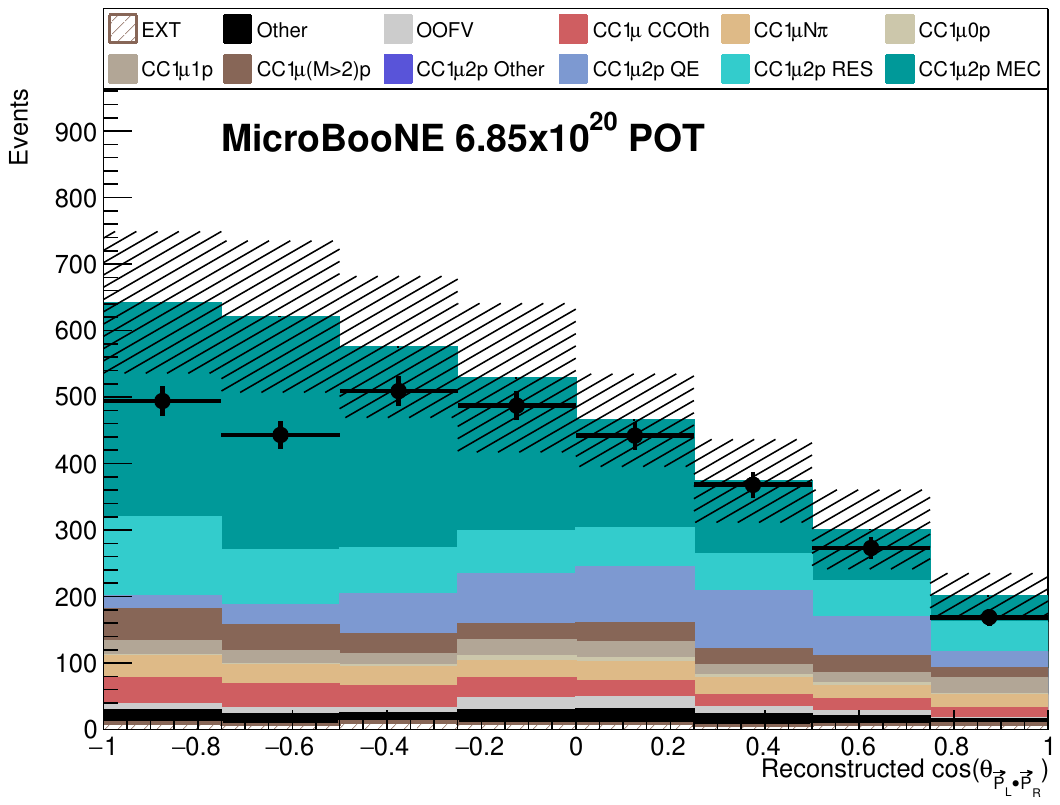}}} \\
\subfloat[][(b)]{{\includegraphics[width=0.8\columnwidth]{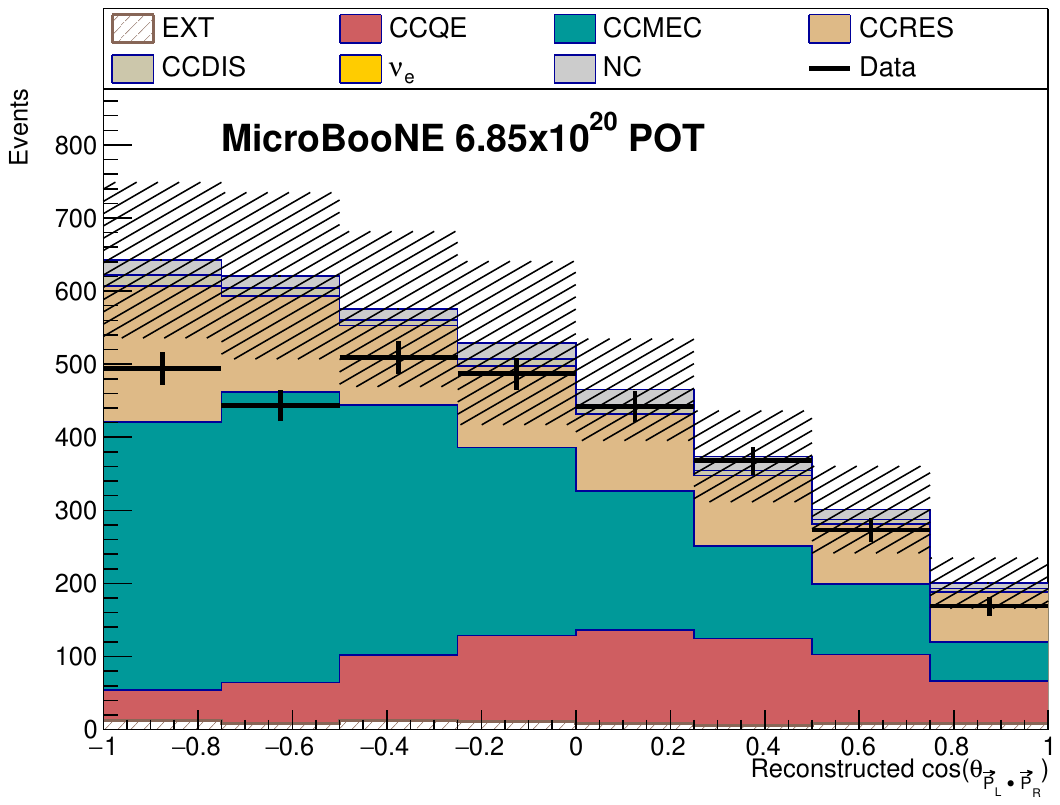}}}
\caption{Yield for the cosine of the opening angle between the protons in the lab frame, $\theta_{\vec{P}_L \cdot \vec{P}_R}$. The selected MC events are broken down into (a) final-state topologies and (b) $\nu$ interaction channels based on the MicroBooNE Tune \cite{uboone_tune} truth information. The error on the data represents the data statistical uncertainty and the hatched region includes both the MC statistical uncertainty of the prediction and systematic uncertainty.
%Note that the Out of FV (grey band) in both plots represents simulated events originating outside of the FV but for which the reconstructed neutrino vertex fell inside of the FV.
Most subsamples are defined in the main text. `OOFV' refers to interactions outside the fiducial volume. `CC1$\mu$(M$>$2)p' events are CC $\nu_\mu$ interactions with more than two outgoing protons above threshold. `CC1$\mu$Oth' are any CC $\nu_\mu$ interactions not included in other samples. `Other' contains any interactions not included in another sample. `CCRES' are resonant CC $\nu_\mu$ interactions and `CCDIS' are deep inelastic scattering CC $\nu_\mu$ interactions.
}
\label{fig:Runall_opening_protons} 
\end{figure}

Distributions of selected events as a function of the cosine of $\theta_{\vec{P}_L \cdot \vec{P}_R}$ can be found in Fig.~\ref{fig:Runall_opening_protons}. The colored bands in each histogram represent events selected from the MC subdivided into different final state topologies [Fig.~\ref{fig:Runall_opening_protons}(a)] and different interaction modes based on the GENIE MicroBooNE Tune prediction [Fig.~\ref{fig:Runall_opening_protons}(b)]. The error bars on the data points are the data statistical uncertainty while the dashed lines represent the uncertainty of the prediction. This uncertainty includes both the statistical uncertainty of the prediction and systematic uncertainty, which dominates over the data statistical uncertainty for this measurement. Contributions from flux modeling and protons-on-target (POT) counting \cite{microboone_flux_prediction}, cross section modeling \cite{uboone_tune}, re-interaction modeling \cite{uboone_reint}, and detector modeling \cite{uboone_detvar} are considered in the calculation of the systematic uncertainties, which was performed using the multiverse techniques described in Section V of Ref.~\cite{uboone_tune}. Uncertainty on the modeling of dirt events is also considered in the systematic uncertainty \cite{uboone_dirt}. We find that our CC1$\mu$2p0$\pi$ signal, represented by the blue-green bands in Fig.~\ref{fig:Runall_opening_protons}(a), constitutes the majority of the prediction. We also find that the CCMEC process, represented by the blue-green band in Fig.~\ref{fig:Runall_opening_protons}(b), has fewer events in the region of $\cos(\theta_{\vec{P}_L \cdot \vec{P}_R}) \simeq 0$ compared to the signal. This is expected as MEC is not the only contributor to CC1$\mu$2p0$\pi$ topologies. 
%The comparison of the data points to the prediction shows reasonable shape agreement as well as normalization agreement.
%Similar agreement is seen in other kinematic variables studied in this analysis~\cite{my_thesis}.

\section{Cross Section Extraction}
\label{xsecextraction}
%%%%%%%%%%%%%%%%%%%%%%%%%%%%%%%%%%% 
Due to detector resolution, efficiency, and smearing effects, our reconstructed kinematic variables require unfolding to obtain the underlying physics quantities. We use the Wiener-SVD technique~\cite{Tang:2017rob} as implemented in \cite{Gardiner:2024gdy}.  In this technique, any bias due to regularization introduced by the unfolding is encoded in a regularization matrix $A_C$ which is applied to all cross-section predictions included in this work. The $A_C$ matrix should be applied to any independent theoretical prediction compared to the data reported in this paper. The data release, unfolded covariance matrices, and $A_C$ matrices are provided in the supplemental material.

%To compare a theoretical prediction to the measurements, the prediction should first be multiplied by $A_C$; this transforms the prediction into the ``regularized space" of the data so that a proper comparison can be made. 

%We have reported the relevant $A_C$ matrices in the Supplemental Materials \cite{supplemental}.
%We use the D'Agostini iterative Bayesian unfolding procedure \cite{agostini_CERN}, as implemented in the RooUnfold software framework \cite{agostini_roounfold}, for this purpose. We determine the optimal number of iterations using the ``L-curve'' technique described in Ref.~\cite{t2k_bayes} and in section 8.3.1 of Ref.~\cite{my_thesis}; briefly, this technique optimizes the smoothness of the solution while also minimizing the number of iterations required. We use two unfolding iterations for the distribution of $\cos(\theta_{\vec{P}_L \cdot \vec{P}_R})$, four unfolding iterations for the distribution of $\cos(\theta_{\vec{P}_{\mu} \cdot \vec{P}_{\mathrm{sum}}})$, and two unfolding iterations for the distribution of $\delta P_T$. The uncertainty associated with the unfolding is determined by comparing the result of a single iteration and that of the total number of iterations (two or four).  
Unfolded distributions are then normalized by the number of target nuclei and the total integrated neutrino flux to produce a cross section. 
We validate the Wiener-SVD unfolding technique by performing fake data studies before unfolding the selected BNB data events.

\begin{figure}[!h]
\captionsetup[subfigure]{labelformat=empty}  
\centering
\subfloat[][(a)]{\includegraphics[width=0.8\columnwidth]{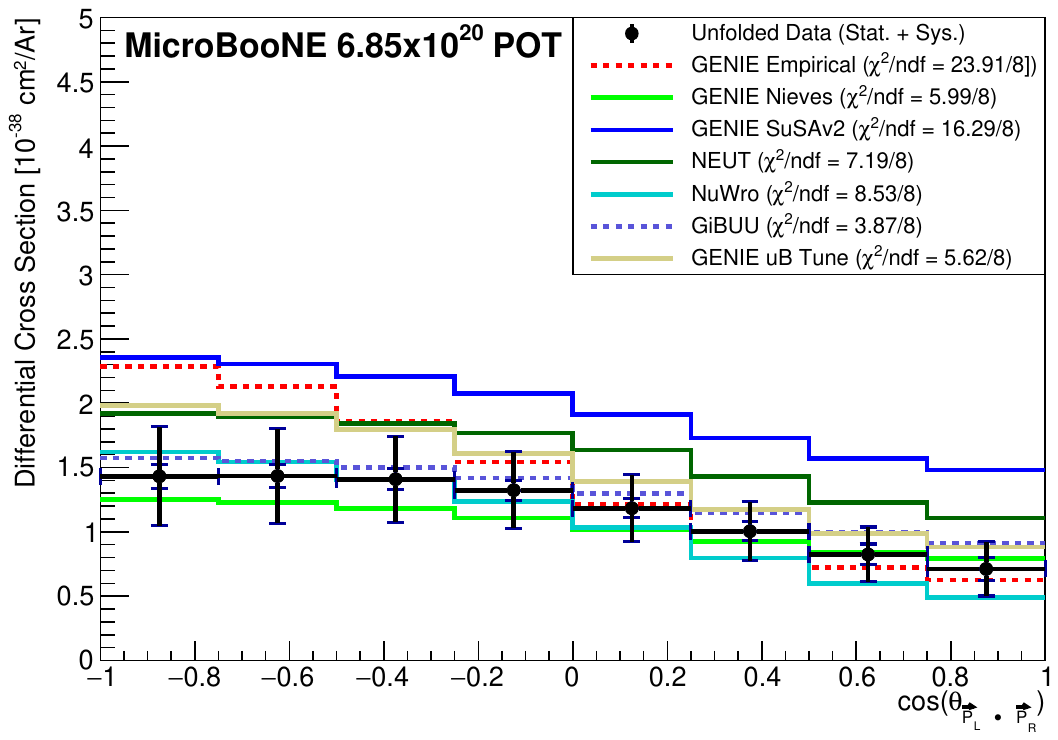}\label{protons}} \\
\subfloat[][(b)]{\includegraphics[width=0.8\columnwidth]{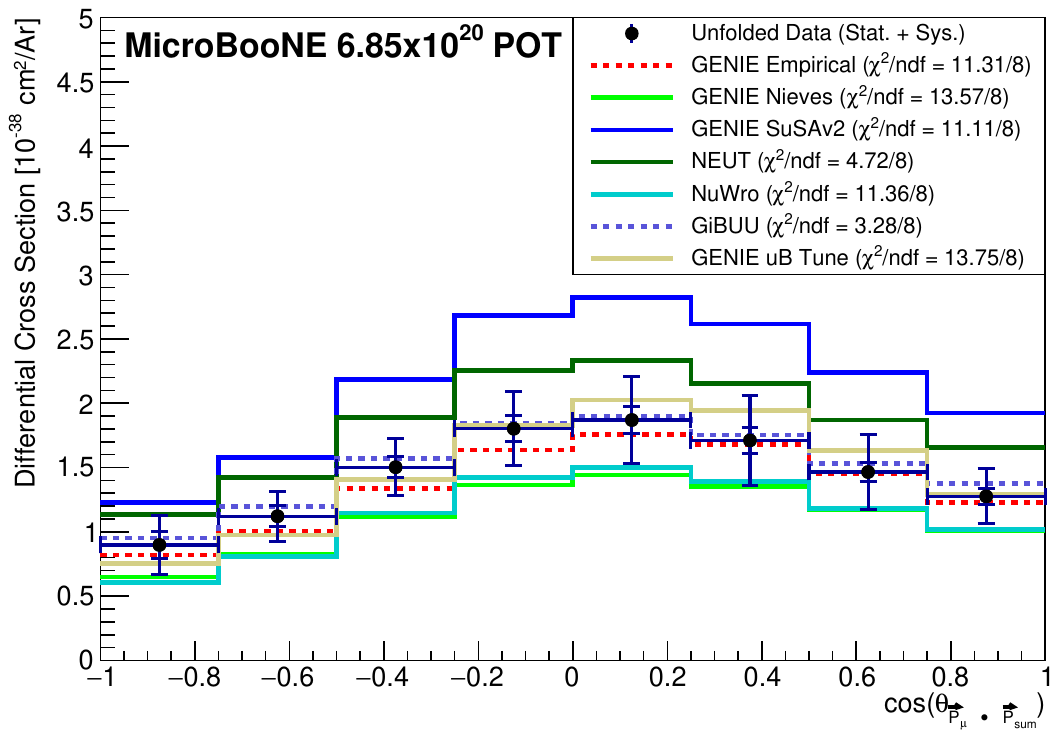}\label{mu_both}}
\caption{Single differential cross sections as a function of (a) the cosine of the opening angle between the protons in the lab frame, $\cos(\theta_{\vec{P}_L \cdot \vec{P}_R})$, and (b) the cosine of the opening angle between the muon and total proton momentum vector, $\cos(\theta_{\vec{P}_{\mu} \cdot \vec{P}_{\mathrm{sum}}})$. The inner error bands on the data represent the statistical uncertainty while the outer error bands represent the systematic uncertainty. A $\chi^2/\textrm{ndf}$, considering systematic and statistical uncertainties, is calculated between the data and each model set curve. Details of the generator predictions are given in \autoref{EventSelection}.}
\label{fig:xsec_opening_angles}
\end{figure}

Fig.~\ref{fig:xsec_opening_angles} shows the single differential cross sections as functions of cosine of $\theta_{\vec{P}_L \cdot \vec{P}_R}$ [Fig.~\ref{fig:xsec_opening_angles}(a)] and cosine of $\theta_{\vec{P}_{\mu} \cdot \vec{P}_{\mathrm{sum}}}$ [Fig.~\ref{fig:xsec_opening_angles}(b)]. The black points represent the extracted cross section from data with the inner error bands representing the statistical uncertainty and the outer error bands representing the systematic uncertainty. 
%The systematic uncertainties include both an overall normalization contribution (9.3\%) and also a point-to-point (``shape'') contribution (11.4\%), while the statistical uncertainty is 4.5\%. 
%Although MEC modeling uncertainties enter predominantly only through the efficiency, they are still the dominant cross-section uncertainties in some bins due to the large uncertainties assigned to this model which is itself due to the lack of previous measurements. 
The dominant uncertainties on the total event rate prediction are comprised of the systematic uncertainties on modelling MEC interactions (which enter primarily through their impact on the signal selection efficiency) and uncertainties on modelling the detector response. The fractional uncertainty in each bin ranges between 13\% and 24\%.

%In addition to the six sources of systematic uncertainty mentioned above, an uncertainty on the choice of the number of unfolding iterations is also included in the systematic error band. 
%The cross sections for the GENIE Empirical, GENIE Nieves, GENIE SuSAv2, NuWro, NEUT and GiBUU model sets were derived using methods described in Ref.~\cite{marley}. The GENIE MicroBooNE Tune cross-section curve is created by selecting generated CC1$\mu$2p0$\pi$ signal events from the Overlay MC and NuWro samples, respectively. Each distribution was then normalized by the number of target nucleons and the total integrated flux to produce a cross-section curve. 
%using the total covariance matrices found in the Supplemental Materials \cite{supplemental}.

%Fake data tests, using NuWro MC and GENIE MicroBooNE Tune with twice the MEC contribution, were performed to test the robustness of the analysis, and no significant biases were found.

When extracting the cross section from data, we find that the GiBUU, GENIE MicroBooNE Tune, and GENIE Nieves models are compatible with our data in $\cos(\theta_{\vec{P}_L \cdot \vec{P}_R})$, and the GiBUU and NEUT models have the best agreement with our data in $\cos(\theta_{\vec{P}_{\mu} \cdot \vec{P}_{\mathrm{sum}}})$. 
%We also find that the GENIE Nieves and NuWro models have the best agreement with our data across the full ranges of $\cos(\theta_{\vec{P}_L \cdot \vec{P}_R})$ and $\cos(\theta_{\vec{P}_{\mu} \cdot \vec{P}_{\mathrm{sum}}})$. 
The $\chi^2$ per degree of freedom ($\chi^2/\textrm{dof}$) is calculated between the data and each model set curve. The number of degrees of freedom is equal to the number of histogram bins. Both systematic and statistical uncertainties on the data are considered in this calculation.

%The GENIE MicroBooNE Tune and GENIE Empirical models tend to predict higher cross sections than indicated by our data in regions of low $\cos(\theta_{\vec{P}_L \cdot \vec{P}_R})$ and high $\cos(\theta_{\vec{P}_{\mu} \cdot \vec{P}_{\mathrm{sum}}})$. Although the GENIE MicroBooNE Tune uses all the same model elements as the GENIE Nieves sample, we find an overall difference in shape and normalization between these two curves. This is most likely caused by the increased CC2p2h cross section normalization and interpolated CC2p2h cross section shape that are utilized in the GENIE MicroBooNE Tune. Additionally, the GENIE SuSAv2 model is overpredicted in both distributions which is likely caused by the increased CC2p2h cross section normalization.

The GENIE Empirical model over-predicts the normalization in the region of low $\cos(\theta_{\vec{P}_L \cdot \vec{P}_R})$, and has the worst agreement with data as evidenced by the $\chi^2/\textrm{dof} = 23.91/7$. The GENIE SuSAv2 model is over-predicted throughout the reported phase space in both opening-angle distributions which is likely caused by the increased CC2p2h cross-section normalization. Despite the tune of CC2p2h cross-section parameters, the GENIE MicroBooNE and GENIE Nieves predictions have approximately equal agreement with data in both opening angle distributions. The GiBUU model set is found to have the best agreement in both distributions. 

\begin{figure}[!h]
\includegraphics[width=1.0\columnwidth]{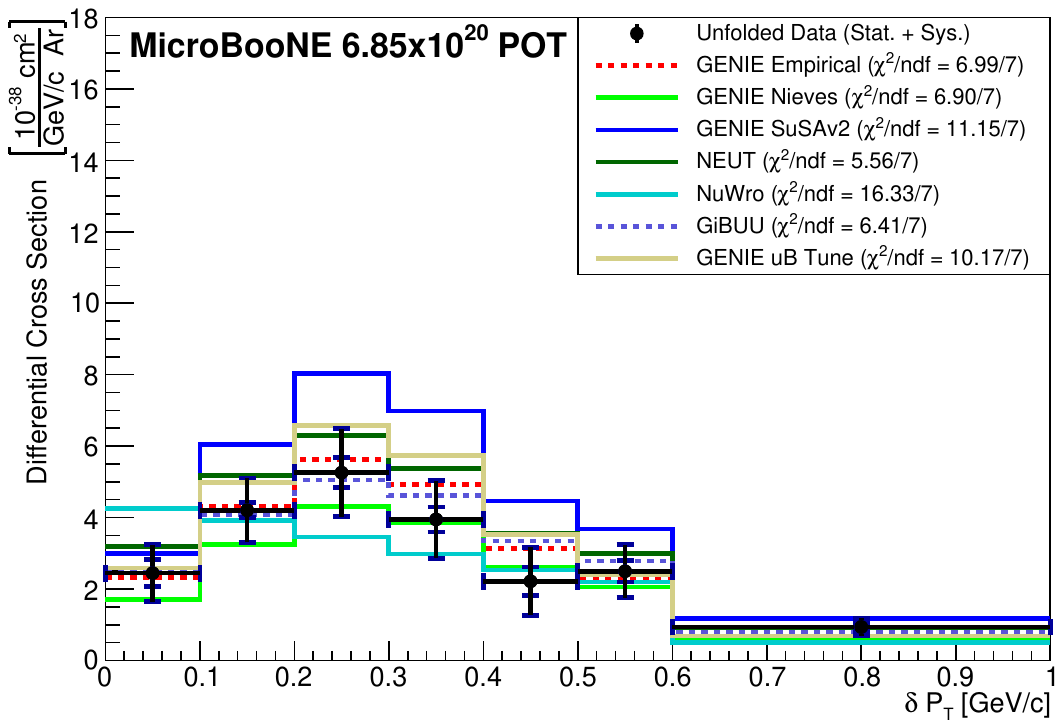}
\caption{Single differential cross section as a function of the magnitude of the transverse momentum of the final state system, $\delta P_T = \left|\vec{P}^{\mu}_T + \vec{P}^L_T + \vec{P}^R_T\right|$. The inner error bands on the data represent the statistical uncertainty while the outer error bands represent the systematic uncertainty. A $\chi^2/\textrm{ndf}$, considering systematic and statistical uncertainties, is calculated between the data and each model set curve. Details of the generator predictions are given in \autoref{EventSelection}.}
\label{fig:xsec_delta_PT}
\end{figure}

We further show the single differential cross sections for the data and the models as functions of $\delta P_T$ in Fig.~\ref{fig:xsec_delta_PT}. In this variable, NuWro has a larger normalization than any other model or data in the first bin which drives the high $\chi^{2}/\textrm{dof} = 16.33/6$. The reason for this is that Nieves and NuWro form their initial hadronic states in different ways. In the GENIE implementation of the GENIE Nieves model set, two nucleons are selected from the Fermi sea of the nucleus \cite{mec}. The momentum of each nucleon is then randomly sampled from a distribution of the initial state nucleon momentum \cite{nieves_mec} formed from the LFG nuclear model \cite{LFG}. In NuWro, the selection of the two nucleons and their momenta is similar to the GENIE implementation of GENIE Nieves, but the two nucleons are required to have back-to-back momenta in the initial state \cite{nuwro_mec}. The over-prediction of NuWro at low $\delta P_T$ has also been observed in Ref.~\cite{lars}, which also finds this over-prediction in the absence of FSI, indicating that the excess is an initial state effect. 
%Our data indicates that there is a preference for the NEUT model set. 
Our data shows agreement with the NEUT, GiBUU, GENIE Nieves, and GENIE Empirical model sets.
%Cross section values for different kinematic variables can be found in the Supplemental Material \cite{supplemental}.
The GENIE SuSAv2 model has an over-predicted normalization, whilst the GENIE MicroBooNE tune prediction has a larger disagreement with the data than its base model, GENIE Nieves.

%\begin{figure}
%\includegraphics[width=1.0\columnwidth]{Figures/xsec_delta_PT.pdf}
%\caption{TEST TO SEE IF A FOURTH FIGURE WILL FIT.}
%\label{fig:TEST}
%\end{figure}

\section{Conclusion}
\label{Conclusion}

In this letter, we present the first measurement of single differential cross sections of CC1$\mu$2p0$\pi$ events on argon. Events containing exactly one muon, two protons, and no other mesons are selected from BNB data. We extract differential cross sections as functions of three kinematic variables, $\cos(\theta_{\vec{P}_L \cdot \vec{P}_R})$, $\cos(\theta_{\vec{P}_{\mu} \cdot \vec{P}_{\mathrm{sum}}})$ and $\delta P_T$, which are found to be sensitive to the formation of 2p2h pairs through MEC and FSI processes. We compare our extracted cross sections to those predicted by the GENIE, GiBUU, NEUT, and NuWro generators, including four different GENIE model sets. These cross-section models span a range of nuclear models, QE and MEC models, and hadron transport models. We find that the GiBUU prediction shows the best overall shape agreement in the kinematic variables describing opening angles. The $\delta P_T$ variable is sensitive to the different initial hadronic states, and we find that the NEUT prediction gives the best overall description of the production of CC1$\mu$2p0$\pi$ final states. 
%None of the models used for comparisons in this work include SRCs in the nuclear model; their inclusion could further improve the agreement with our measured differential cross sections. 
This is the first differential cross-section measurement of two-proton final states, and therefore the first time 2p2h/MEC models have been compared to data in an analysis dominated by such interactions. This provides valuable input for future model development toward precision neutrino physics measurements. In addition, these measured CC1$\mu$2p0$\pi$ cross sections can be used to reinterpret data from existing experiments that cannot distinguish 2p2h final states from other CC interaction mechanisms.

%ACKNOWLEDGEMENTS
%%%%%%%%%%%%%%%%%%%%%%%%%%%%%%%%%%%%%%%%%%%%
This document was prepared by the MicroBooNE collaboration using the
resources of the Fermi National Accelerator Laboratory (Fermilab), a
U.S. Department of Energy, Office of Science, Office of High Energy Physics HEP User Facility.
Fermilab is managed by Fermi Forward Discovery Group, LLC, acting
under Contract No. 89243024CSC000002.  MicroBooNE is supported by the
following: 
the U.S. Department of Energy, Office of Science, Offices of High Energy Physics and Nuclear Physics; 
the U.S. National Science Foundation; 
the Swiss National Science Foundation; 
the Science and Technology Facilities Council (STFC), part of the United Kingdom Research and Innovation; 
the Royal Society (United Kingdom); 
the UK Research and Innovation (UKRI) Future Leaders Fellowship; 
and the NSF AI Institute for Artificial Intelligence and Fundamental Interactions. 
Additional support for 
the laser calibration system and cosmic ray tagger was provided by the 
Albert Einstein Center for Fundamental Physics, Bern, Switzerland. We 
also acknowledge the contributions of technical and scientific staff 
to the design, construction, and operation of the MicroBooNE detector 
as well as the contributions of past collaborators to the development 
of MicroBooNE analyses, without whom this work would not have been 
possible. 
For the purpose of open access, the authors have applied 
a Creative Commons Attribution (CC BY) public copyright license to 
any Author Accepted Manuscript version arising from this submission.

%BIBLIOGRAPHY
%%%%%%%%%%%%%%%%%%%%%%%%%%%%%%%%%%%%%%%%%
\bibliographystyle{unsrtnat}
\bibliography{CC2P}

\end{document}